# Chromo-field flux sheets as confining gauge field configurations in the SU(N) Euclidean Yang-Mills theory in the Landau gauge

Hiroshi Matsuoka

Department of Physics, Illinois State University, Normal, Illinois, 61790-4560, U.S.A.

**Abstract**

For the four-dimensional SU(N) Euclidean Yang-Mills theory in the Landau gauge, we present two sets of gauge field configurations that satisfy the Euclidean equations of motion. These configurations generate four-dimensional chromo-field flux sheets whose spatial cross sections are three-dimensional chromo-field flux tubes. In lattice simulations, they may be detected as center vortices. The first set of gauge field configurations generates chromo-electric flux tubes that should contribute to a chromo-electric flux tube between two static color charges. The string tension $\sigma_r$ for two static color charges in representation $r$ then naturally satisfies the Casimir scaling. Applying a gauge transformation to this set of gauge field configurations, we can transform them into those in the maximal Abelian gauge. These transformed configurations generate chromo-electric flux tubes that should contribute to those observed between two static quarks in lattice simulations performed in the maximal Abelian gauge. The second set of gauge field configurations generates chromo-magnetic flux tubes. When rotated in a plane that includes the temporal $x_4$-axis and is perpendicular to the flux tube axis, the rotated gauge field configuration generates a chromo-electric flux tube and should contribute to the chromo-electric flux tubes observed in lattice simulations in the Landau gauge. We also argue that when regulated on a lattice, any of the flux sheet gauge field configuration with a finite flux sheet thickness is located on the Gribov horizon in the infinite lattice volume limit. We thus suggest that these sets of gauge field configurations contribute significantly to the low energy properties of QCD, particularly the quark confinement.

## I. INTRODUCTION AND A SUMMARY OF OUR RESULTS

The confinement of quarks and gluons within color singlet hadrons is one of the most important low energy properties of quantum chromodynamics (QCD). Its precise mechanism, however, remains unresolved [1]. In this article, we present two sets of gauge field configurations for the four-dimensional SU(N) Euclidean Yang-Mills theory in the Landau gauge and suggest that they contribute significantly to the low-energy properties of QCD, particularly the quark confinement.

Each of these gauge field configurations is built from two non-vanishing components of the gauge field, a diagonal component, $A_{34}$ or $\mathbf{A}_3$, where 3 refers to the diagonal generator of the Lie algebra for SU(N), and an off-diagonal component $\mathbf{A}_2$. For higher values of *N*, there are more diagonal generators so that there are more of these types of configurations, which may explain some differences in the low energy properties of the Yang-Mills theories with different values of *N* (*e.g.*, SU(2) versus SU(3)).

We will construct these gauge field configurations by first assuming that they do not depend on the temporal coordinate $x_4$. However, as the Euclidean Lagrangian is symmetric with respect to rotations in four-dimensional Euclidean space, these configuration can be always rotated in the Euclidean space to become those that depend on $x_4$.

These gauge field configurations are then required to satisfy the Euclidean equations of motion so that in the configurational space for the gauge field, they correspond to stationary points, which implies that in a neighborhood of each of these configurations, there are many other configurations whose values for the Euclidean action $S_\mathrm{E}$ for the Yang-Mills theory are close to that of the configuration of interest.

We will show that these gauge field configurations generate four-dimensional chromo-field flux sheets whose spatial cross sections are three-dimensional chromo-field flux tubes. However, at the core of a chromo-field flux tube, the non-zero components of the gauge field diverge logarithmically so that we must regularize these divergences by replacing the Euclidean path integral for the partition function of the Yang Mills theory by the corresponding lattice gauge theory on a four-dimensional lattice. As the non-zero components of the gauge field tend to diverge near the center of a chromo-field flux tube, the link variables corresponding to these gauge field components start to sample non-trivial center elements in the gauge group so that in lattice simulations, these chromo-field flux tubes may be detected as center vortices.

When simulated on a finite lattice, these gauge field configurations can have arbitrarily small values of the Euclidean action $S_E$ so that they should contribute significantly to the vacuum state of the Yang-Mills theory with high statistical weights as the statistical weight or probability for each configuration is proportional to $\exp(-S_E)$. This is why we believe that these gauge field configurations must significantly contribute to the low-energy properties of the SU(N) Yang Mills theory, particularly QCD.

One set of gauge field configurations that will be discussed in Sec.III.B generates chromo-electric flux tubes that should contribute to a chromo-electric flux tube between two static quarks. The string tension $\sigma_r$ for two static color charges in representation $r$ then naturally satisfies the Casimir scaling. By applying a gauge transformation to these configurations, we can transform them into those in the maximal Abelian gauge. We will find these transformed configurations to generate chromo-electric flux tubes, where chromo-electric fields are related to the Euclidean magnetic current densities induced by the transformed gauge fields in a way similar to the way the Abelian projected chromo-electric field is related to the monopole current density as observed in lattice simulations [2].

Another set of gauge field configurations that will be discussed in Sec.III.C generates chromo-magnetic flux tubes. When rotated in a plane that includes the temporal $x_4$-axis and is perpendicular to the flux tube axis, each of these configurations acquires non-zero chromo-electric field along the flux tube axis so that the rotated gauge field configuration generates a chromo-electric flux tube. Such chromo-electric flux tubes have been observed in lattice simulations [3].

The way the chromo-field flux tubes are generated is somewhat analogous to the way magnetic vortices are formed in type II superconductors. The off-diagonal component $\mathbf{A}_2$ of the gauge field acquires a constant non-zero magnitude $\sqrt{\mathbf{A}_2 \cdot \mathbf{A}_2} = |A_2^\infty|$ throughout the Euclidean space except where the diagonal component, $A_{34}$ or $\mathbf{A}_3$, of the gauge field takes non-zero values so that the chromo-fields controlled by the diagonal component are squeezed into a flux tube. The off-diagonal component thus plays a role of the Cooper pair condensate while the diagonal component plays a role of the vector potential for the magnetic field in the magnetic vortex. We will find the thickness $\lambda$ of the flux tubes to be inversely proportional to $|A_2^\infty|$ as $\lambda \equiv 1/(gC_{123}|A_2^\infty|)$, where $g$ is the coupling constant of the Yang Mills theory and $C_{123}$ is the structure constant of the Lie algebra for SU(N).

It is interesting to note that in lattice simulations, the thickness of average chromo-electric flux tubes for the SU(3) lattice gauge theory has been estimated to be $\lambda_{\text{lattice}} \sim 0.1$ fm $\sim (2 \text{ GeV})^{-1}$ [4] while the mass-dimension 2 condensate has been estimated to be also $\sqrt{g^2 \langle A^{\mu a} A_\mu^a \rangle} \sim 2$ GeV [5], which appears to be consistent with $\lambda \equiv 1/(gC_{123}|A_2^\infty|)$. In addition, the lowest-lying glueball mass has been also estimated to be $m_{0++} \sim 2$ GeV [6] so that $\lambda_{\text{lattice}} \sim 1/\sqrt{g^2 \langle A^{\mu a} A_\mu^a \rangle} \sim 1/m_{0++}$.

We may also suggest that these gauge field configurations disorder the vacuum state of the SU(N) Euclidean Yang Mills theory producing the area law for the expectation value for the Wilson loop and generating a finite correlation length or a mass gap for the Euclidean correlation functions for gauge invariant quantities.

It has been also proposed [7] that in the Landau gauge, the gauge field configurations located near or on the Gribov horizon, the boundary of the Gribov region [8], are responsible for the low energy properties of the SU(N) Yang-Mills theories. In Sec.III.D, we will argue that when regulated on a lattice with a volume $V = L^4$, a flux sheet gauge field configuration belonging to one of the two sets to be presented in Sec.III.B and Sec.III.C is located on the Gribov horizon as long as its flux sheet thickness, $\lambda = 1/(gC_{123}|A_2^\infty|)$, is much shorter than $L$ (i.e., $\lambda \ll L$) so that in the infinite lattice volume limit, any flux sheet gauge field configuration with a finite flux sheet thickness is located on the Gribov horizon.

This article is organized as follows. In Sec.II, we introduce the partition function for the SU(N) Euclidean Yang Mills theory, the Euclidean equations of motion, and the Euclidean Bianchi identity for the Euclidean chromo-fields. In Sec.III, we present two sets of gauge field configurations that satisfy the Euclidean equations of motion and generate four-dimensional chromo-field flux sheets whose spatial cross sections are three-dimensional chromo-field flux tubes. In Sec.III.B, we discuss the set of gauge field configurations that generate chromo-electric flux tubes while in Sec.III.C, we discuss the set of gauge field configurations that generate chromo-magnetic flux tubes. In Sec.IV, we close the article with some conclusions and outlook.

## II. THE SU(N) EUCLIDEAN YANG MILLS THEORY

### A. The partition function and the Euclidean action

We can express the partition function for the SU(N) Yang Mills theory at a finite inverse temperature $\beta$ as a path integral with the Euclidean action $S_E$:

$$Z = \mathsf{N} \int \left( D\mathbf{A}_a^{(E)} \right)\left( DA_{a4}^{(E)} \right) e^{-S_E}, \tag{2.1}$$

where $\mathsf{N}$ is a numerical constant. $\mathbf{A}_a^{(E)}$ and $A_{a4}^{(E)}$ are the spatial and the "temporal" components of the Euclidean gauge field $A_{a\mu}^{(E)}$, where $a$ is the color index for the gauge field. The Euclidean action $S_E$ is related to the Euclidean Lagrangian $L_E$ by

$$S_E = \int_0^\beta dx_4^{(E)} \int d\mathbf{r} L_E \tag{2.2}$$

and $L_E$ is defined by

$$L_{\text{E}} \equiv \frac{1}{4} F^{(E)}_{a\mu\nu} F^{(E)}_{a\mu\nu} = \frac{1}{2}\left(\mathbf{E}^{(E)}_a \cdot \mathbf{E}^{(E)}_a + \mathbf{B}^{(E)}_a \cdot \mathbf{B}^{(E)}_a\right), \qquad (2.3)$$

where the Euclidean field strength $F^{(E)}_{a\mu\nu}$ is defined by

$$F^{(E)}_{a\mu\nu} \equiv \partial^{(E)}_\mu A^{(E)}_{a\nu} - \partial^{(E)}_\nu A^{(E)}_{a\mu} - g C_{abc} A^{(E)}_{b\mu} A^{(E)}_{c\nu}, \qquad (2.4)$$

where $g$ is the coupling constant of the Yang Mills theory and $C_{123}$ is the structure constant of the Lie algebra for SU(N). The Euclidean chromo-electric and chromo-magnetic fields are then defined by

$$E^{(E)}_{ak} \equiv F^{(E)}_{ak4} = \partial^{(E)}_k A^{(E)}_{a4} - \partial^{(E)}_4 A^{(E)}_{ak} - g C_{abc} A^{(E)}_{bk} A^{(E)}_{c4} \qquad (2.5)$$

and

$$B^{(E)}_{ak} \equiv \frac{1}{2} \varepsilon^{kij} F^{(E)}_{aij} = \left(\nabla^{(E)} \times \mathbf{A}^{(E)}_a - \frac{1}{2} g C_{abc} \mathbf{A}^{(E)}_b \times \mathbf{A}^{(E)}_c\right)_k. \qquad (2.6)$$

For the rest of this article, we will drop the superscript "(E)" and assume that the temperature is set to be zero so that $\beta = \infty$.

### B. The Euclidean equations of motion

We can express the Euler-Lagrange equation for the Euclidean gauge field as

$$\partial_\mu F_{a\mu\nu} = J_{a\nu} \equiv g C_{abc} A_{b\mu} F_{c\mu\nu}, \qquad (2.7)$$

where we have defined the Euclidean color current density $J_{a\nu}$. This equation can be written as the Euclidean equations of motion for the Euclidean chromo-electric and chromo-magnetic fields:

$$\nabla \cdot \mathbf{E}_a = J_{a4} \tag{2.8}$$

and

$$\nabla \times \mathbf{B}_a + \partial_4 \mathbf{E}_a = -\mathbf{J}_a, \tag{2.9}$$

where

$$J_{a4} = gC_{abc}\mathbf{A}_b \cdot \mathbf{E}_c \tag{2.10}$$

and

$$-\mathbf{J}_a = gC_{abc}\left(A_{b4}\mathbf{E}_c + \mathbf{A}_b \times \mathbf{B}_c\right). \tag{2.11}$$

### C. The Euclidean Bianchi identity

The Euclidean dual field strength is defined by

$$\tilde{F}_{a\mu\nu} \equiv \frac{1}{2} \varepsilon^{\mu\nu\alpha\beta} F_{a\alpha\beta} \tag{2.12}$$

and satisfies the following Euclidean Bianchi identity:

$$\partial_\mu \tilde{F}_{a\mu\nu} = K_{a\nu} \equiv gC_{abc}A_{b\mu}\tilde{F}_{c\mu\nu}, \tag{2.13}$$

where the Euclidean magnetic current density $K_{a\nu}$ is defined. As the Euclidean dual field strength is related to the Euclidean chromo-electric and chromo-magnetic fields by $\tilde{F}_{aij} = \varepsilon^{ijk} E_{ak}$ and $\tilde{F}_{ai4} = B_{ai}$, the Euclidean Bianchi identity becomes

$$\nabla \cdot \mathbf{B}_a = K_{a4} \tag{2.14}$$

and

$$-\nabla \times \mathbf{E}_a - \partial_4 \mathbf{B}_a = \mathbf{K}_a, \tag{2.15}$$

where

$$K_{a4} = gC_{abc}\mathbf{A}_b \cdot \mathbf{B}_c = -\frac{1}{2}gC_{abc}\nabla \cdot (\mathbf{A}_b \times \mathbf{A}_c) \tag{2.16}$$

and

$$\mathbf{K}_a = -gC_{abc}(A_{b4}\mathbf{B}_c + \mathbf{A}_b \times \mathbf{E}_c) = gC_{abc}\left\{\nabla \times (\mathbf{A}_b A_{c4}) + \frac{1}{2}\partial_4(\mathbf{A}_b \times \mathbf{A}_c)\right\} \qquad (2.17)$$

## III. THE GAUGE FIELD CONFIGURATIONS WITH CHROMO-FIELD FLUX TUBES

### A. The rotational symmetry of the Euclidean Lagrangian

In the next two sections, we present two sets of gauge field configurations that satisfy both the Euclidean equations of motion and the Euclidean Bianchi identity but do not depend on the temporal coordinate $x_4$. As the Euclidean Lagrangian is symmetric with respect to rotations in the four-dimensional Euclidean space, these gauge field configuration can be always rotated in the Euclidean space to become those that depend on $x_4$.

### B. The gauge configuration with a chromo-electric flux tube

#### 1. The gauge field configuration and the chromo-fields

Consider a gauge field configuration that satisfies

$$\mathbf{A}_2 = A_2(r)\hat{z} \qquad (3.1)$$

and

$$A_{34} = A_{34}(r), \qquad (3.2)$$

where we use the cylindrical coordinate system, where at each spatial location $\mathbf{r} = (r,\phi,z)$, we have three unit vectors, $\hat{r}$, $\hat{\phi}$, and $\hat{z}$. All the other components of the gauge field are zero. This gauge field configuration satisfies both $\nabla \cdot \mathbf{A}_a = 0$ and $\partial_4 A_{a4} = 0$ so that it satisfies the Landau gauge condition, $\nabla \cdot \mathbf{A}_a + \partial_4 A_{a4} = 0$.

We then find

$$\mathbf{E}_1 = -gC_{123}A_{34}\mathbf{A}_2 = -gC_{123}A_{34}(r)A_2(r)\hat{z}, \tag{3.3}$$

$$\mathbf{E}_3 = \nabla A_{34} = \frac{dA_{34}}{dr}\hat{r}, \tag{3.4}$$

and

$$\mathbf{B}_2 = \nabla \times \mathbf{A}_2 = -\frac{dA_2}{dr}\hat{\phi}, \tag{3.5}$$

while $\mathbf{E}_2 = 0$ and $\mathbf{B}_1 = \mathbf{B}_3 = 0$. We will show that these fields decay exponentially as $r$ is increased so that they represent a chromo-electric flux sheet whose three-dimensional cross section is a chromo-electric flux tube parallel to the $z$-axis.

## 2. The equations for the gauge field

The equation of motion for $\mathbf{E}_3$, $\nabla \cdot \mathbf{E}_3 = J_{34}$, together with

$$J_{34} \equiv -gC_{123}\mathbf{A}_2 \cdot \mathbf{E}_1 = (gC_{123}A_2)^2 A_{34}, \tag{3.6}$$

becomes the following equation for $A_{34}$

$$\nabla^2 A_{34} = \nabla \cdot \mathbf{E}_3 = J_{34} = (gC_{123}A_2)^2 A_{34} \tag{3.7}$$

so that

$$\left(\frac{d^2}{dr^2} + \frac{1}{r}\frac{\partial}{\partial r}\right)A_{34} = (gC_{123}A_2)^2 A_{34}. \tag{3.8}$$

The equation of motion for $\mathbf{B}_2$, $\nabla \times \mathbf{B}_2 = -\mathbf{J}_2$, together with

$$\mathbf{J}_2 = gC_{123}A_{34}\mathbf{E}_1 = (gC_{123}A_{34})^2 \mathbf{A}_2 \tag{3.9}$$

becomes the following equation for $A_2$

$$\nabla^2 \mathbf{A}_2 = -\nabla \times \nabla \times \mathbf{A}_2 = -\nabla \times \mathbf{B}_2 = \mathbf{J}_2 = (gC_{123}A_{34})^2 \mathbf{A}_2 \qquad (3.10)$$

so that

$$\left(\frac{d^2}{dr^2} + \frac{1}{r}\frac{\partial}{dr}\right) A_2 = (gC_{123}A_{34})^2 A_2. \qquad (3.11)$$

### 3. The solution for the gauge field away from the core of a chromo-field flux tube

To solve the above equations for $A_{34}$ and $A_2$, we impose the following asymptotic boundary conditions:

$$\lim_{r \to \infty} A_{34} = 0 \qquad (3.12)$$

and

$$\lim_{r \to \infty} A_2 = \text{const} = A_2^\infty. \qquad (3.13)$$

For dimensionless variables, $\hat{A}_{34}$ and $\delta\hat{A}_2$, defined by

$$A_{34} \equiv |A_2^\infty| \hat{A}_{34} \qquad (3.14)$$

and

$$A_2 \equiv A_2^\infty \left(1 + \delta\hat{A}_2\right), \qquad (3.15)$$

the above equations become

$$\left(\frac{d^2}{d\rho^2} + \frac{1}{\rho}\frac{d}{d\rho}\right)\hat{A}_{34} = \left(1 + \delta\hat{A}_2\right)^2 \hat{A}_{34} \qquad (3.16)$$

and

$$\left(\frac{d^2}{d\rho^2} + \frac{1}{\rho}\frac{d}{d\rho}\right)\delta\hat{A}_2 = \left(\hat{A}_{34}\right)^2 \left(1 + \delta\hat{A}_2\right), \qquad (3.17)$$

where the scaled radial distance $\rho$ is defined by

$$\rho \equiv g C_{123} |A_2^\infty| r. \tag{3.18}$$

Assuming $|\delta\hat{A}_2| \ll 1$ for $\rho \gg 1$, which we will justify later, we can approximate these equation for $\hat{A}_{34}$ and $\delta\hat{A}_2$ by

$$\left(\frac{d^2}{d\rho^2} + \frac{1}{\rho}\frac{d}{d\rho}\right)\hat{A}_{34} \cong \hat{A}_{34} \tag{3.19}$$

and

$$\left(\frac{d^2}{d\rho^2} + \frac{1}{\rho}\frac{d}{d\rho}\right)\delta\hat{A}_2 \cong (\hat{A}_{34})^2. \tag{3.20}$$

The solution of the equation for $\hat{A}_{34}$ is then

$$\hat{A}_{34} \cong c_4 K_0(\rho), \tag{3.21}$$

where $K_0$ is the modified Bessel function of the second kind and $c_4$ is a constant that sets the scale for $\hat{A}_{34}$. We then obtain

$$A_{34} \cong c_4 |A_2^\infty| K_0(-g C_{123} |A_2^\infty| r), \tag{3.22}$$

The equation for $\delta\hat{A}_2$ then becomes

$$\frac{1}{\rho}\frac{d}{d\rho}\left(\rho\frac{d}{d\rho}\right)\delta\hat{A}_2 \cong (c_4 K_0)^2. \tag{3.23}$$

Since for $\rho \gg 1$,

$$K_\nu(\rho) \cong \sqrt{\frac{\pi}{2\rho}} \exp(-\rho), \tag{3.24}$$

we can further approximate the equation for $\delta\hat{A}_2$ by

$$\frac{d}{d\rho}\left(\rho\frac{d}{d\rho}\right)\delta\hat{A}_2 \cong \left(\frac{\pi}{2}c_4^2\right)\exp(-2\rho). \tag{3.25}$$

By integrating the both sides of this equation, we obtain

$$\rho\frac{d}{d\rho}\delta\hat{A}_2 = \int_\infty^\rho d\rho'\frac{d}{d\rho'}\left(\rho'\frac{d}{d\rho'}\right)\delta\hat{A}_2(\rho') = -\left(\frac{\pi c_4^2}{4}\right)\exp(-2\rho), \tag{3.26}$$

where we have assumed

$$\lim_{\rho\to\infty}\rho\frac{d}{d\rho}\delta\hat{A}_2(\rho) = 0. \tag{3.27}$$

We then find

$$\frac{d}{d\rho}\delta\hat{A}_2(\rho) = -\left(\frac{\pi c_4^2}{4\rho}\right)\exp(-2\rho). \tag{3.28}$$

By integrating the both sides of this equation, we obtain

$$\delta\hat{A}_2 = -\left(\frac{\pi c_4^2}{4}\right)\left\{-\int_{2\rho}^\infty d\rho'\frac{1}{\rho'}\exp(-\rho')\right\} = -\left(\frac{\pi c_4^2}{4}\right)\text{Ei}(-2\rho), \tag{3.29}$$

where we have assumed $\lim_{\rho\to\infty}\delta\hat{A}_2 = 0$ and Ei is the exponential integral. We then finally obtain

13$$A_2 = A_2^\infty \left\{ 1 - \left(\frac{\pi c_4{}^2}{4}\right) \mathrm{Ei}(-2\rho) \right\}. \tag{3.30}$$

For $r \gg 1$, we then find both $\hat{A}_{34}$ and $\delta\hat{A}_2$ to decay exponentially away from the z-axis:

$$A_{34} \cong \sqrt{\frac{\pi}{2}} c_4 |A_2^\infty| \sqrt{\frac{\lambda}{r}} \exp\left(-\frac{r}{\lambda}\right), \tag{3.31}$$

where we have defined the decay length $\lambda$ for $\hat{A}_{34}$ by

$$\lambda \equiv \frac{1}{gC_{123}|A_2^\infty|}, \tag{3.32}$$

and

$$A_2 \cong A_2^\infty \left\{ 1 + \left(\frac{\pi c_4{}^2}{4}\right) \frac{\xi}{r} \exp\left(-\frac{r}{\xi}\right) \right\}, \tag{3.33}$$

where we have defined the decay length $\xi$ for $\delta\hat{A}_2$ by

$$\xi \equiv \frac{1}{2gC_{123}|A_2^\infty|} = \frac{1}{2}\lambda \tag{3.34}$$

and we have also used

$$\mathrm{Ei}(-2\rho) \cong -\frac{1}{2\rho} \exp(-2\rho). \tag{3.35}$$

$\lambda = 2\xi$ then justifies our initial assumption that $|\delta\hat{A}_2| \ll 1$ for $\rho = r/\lambda \gg 1$.

Note that $A_2$ is somewhat analogous to the Cooper pair condensate in the Ginzburg-Landau theory of superconductors while $A_{34}$ is analogous to the vector potential for a magnetic field in a superconductor. $\lambda$ is then analogous to the penetration length while $\xi$ is analogous to the coherence length. As $\lambda/\xi = 2 > 1/\sqrt{2}$, the chromo-electric flux tube is analogous to a magnetic vortex in a type II superconductor.

### 4. The squeezing of the chromo-electric and chromo-magnetic fields

For $r \gg 1$, we can then show that the chromo-electric fields, $\mathbf{E}_1$ and $\mathbf{E}_3$, decay exponentially with the decay length $\lambda$ for $\hat{A}_{34}$ while the chromo-magnetic field $\mathbf{B}_2$ decays exponentially with the decay length $\xi$ for $\delta \hat{A}_2$:

$$\mathbf{E}_1 = -c_4 g C_{123} \left(A_2^\infty\right)^2 K_0\left(-\frac{r}{\lambda}\right) \hat{z} \cong -\sqrt{\frac{\pi}{2}} c_4 g C_{123} \left(A_2^\infty\right)^2 \sqrt{\frac{\lambda}{r}} \exp\left(-\frac{r}{\lambda}\right) \hat{z}, \qquad (3.36)$$

$$\mathbf{E}_3 \cong c_4 g C_{123} \left(A_2^\infty\right)^2 K_1\left(-\frac{r}{\lambda}\right) \hat{r} \cong \sqrt{\frac{\pi}{2}} c_4 g C_{123} \left(A_2^\infty\right)^2 \sqrt{\frac{\lambda}{r}} \exp\left(-\frac{r}{\lambda}\right) \hat{r}, \qquad (3.37)$$

and

$$\mathbf{B}_2 \cong \left(\frac{\pi c_4^2}{4}\right) \frac{A_2^\infty}{r} \exp\left(-\frac{r}{\xi}\right) \hat{\phi}. \qquad (3.38)$$

We can also show that far away from the z-axis so that $\mathbf{A}_2 \cong A_2^\infty \hat{z}$ and $\mathbf{B}_2 \cong 0$, the chromo-electric field $\mathbf{E}_1$ satisfies

$$\nabla^2 \mathbf{E}_1 \cong \left(g C_{123} A_2^\infty\right)^2 \mathbf{E}_1, \qquad (3.39)$$

which is analogous to the equation describing the Meissner effect for the magnetic field in a superconductor. In fact, this equation follows from $\nabla \times \mathbf{E}_1 = -\mathbf{K}_1$, which is analogous to the equation for Ampere's law for the magnetic field in a superconductor, and $\nabla \cdot \mathbf{E}_3 = J_{34}$, where $J_{34} \cong \left(g C_{123} A_2^\infty\right)^2 A_{34}$, which is analogous to the London equation for a superconductor.

$$\nabla^2 \mathbf{E}_1 = \nabla(\nabla \cdot \mathbf{E}_1) - \nabla \times \nabla \times \mathbf{E}_1 = \nabla \times \mathbf{K}_1 \cong -gC_{123}\nabla \times (\mathbf{A}_2 \times \mathbf{E}_3)$$
$$= -gC_{123}\{(\mathbf{E}_3 \cdot \nabla)\mathbf{A}_2 - (\mathbf{A}_2 \cdot \nabla)\mathbf{E}_3 + \mathbf{A}_2(\nabla \cdot \mathbf{E}_3) - \mathbf{E}_3(\nabla \cdot \mathbf{A}_2)\} \quad (3.40)$$
$$\cong -gC_{123}(A_2^\infty \hat{z})(\nabla \cdot \mathbf{E}_3) \cong (gC_{123}A_2^\infty)^2 \mathbf{E}_1,$$

where $-\mathbf{K}_1 \cong gC_{123}(\mathbf{A}_2 \times \mathbf{E}_3)$.

Similarly, we find

$$\nabla^2 \mathbf{E}_3 = (gC_{123}A_2^\infty)^2 \mathbf{E}_3, \qquad (3.41)$$

which follows from $\nabla \cdot \mathbf{E}_3 = J_{34}$ as

$$\nabla^2 \mathbf{E}_3 = \nabla(\nabla \cdot \mathbf{E}_3) - \nabla \times \nabla \times \mathbf{E}_3 = \nabla(\nabla \cdot \mathbf{E}_3) = \nabla J_{34}$$
$$\cong (gC_{123}A_2^\infty)^2 \nabla A_{34} = (gC_{123}A_2^\infty)^2 \mathbf{E}_3. \qquad (3.42)$$

**5. The gauge field diverges logarithmically at the core of a chromo-electric flux tube**

We can show that at the core of the chromo-electric flux tube, the non-zero components of the gauge field satisfying Eq.(3.8) and Eq.(3.11) diverge logarithmically. We can regularize these divergences by approximating the path integral for the partition function by the corresponding lattice gauge theory on a four-dimensional lattice.

As the non-zero components of the gauge field tend to diverge near the center of a chromo-electric flux tube, the link variables corresponding to these gauge field components start to sample non-trivial center elements in the gauge group so that in lattice simulations, these chromo-electric flux tubes may be detected as center vortices.

**6. The Euclidean action for the gauge field configuration**

As $\mathbf{B}_2 \cdot \mathbf{B}_2 \propto c_4^{\,4}(A_2^\infty)^2$ and $\mathbf{E}_1 \cdot \mathbf{E}_1 + \mathbf{E}_3 \cdot \mathbf{E}_3 \propto c_4^{\,2}(A_2^\infty)^3$, the Euclidean action $S_E$ for the lattice gauge field configuration corresponding to our gauge field configuration is also controlled by $c_4$ and $A_2^\infty$. Particularly, the value of $S_E$ can be arbitrarily small so that these lattice gauge field configurations should contribute significantly to the vacuum state of the Yang-Mills theory with high statistical weights as the statistical weight or probability for each configuration is proportional to $\exp(-S_E)$. This is why we believe that these gauge field configurations must significantly contribute to the low-energy properties of the SU(N) Yang Mills theory, particularly QCD.

### 7. The gauge configuration in the maximal Abelian gauge

In lattice simulations, the chromo-electric flux tube between two static quarks has been observed in the maximal Abelian gauge. To transform our gauge field configuration to those in the maximal Abelian gauge, we apply the following gauge transformation to our gauge field configuration near $z = 0$ to reduce the size of the off-diagonal component $\mathbf{A}_2$, to which $A_2^\infty$ contributes the most. For simplicity, we will work with the gauge group SU(2) so that $C_{123} = \varepsilon^{123} = 1$ and the corresponding generators are $\sigma_i/2$ ($i = 1, 2, 3$), where $\sigma_1$, $\sigma_2$, and $\sigma_3$ are the Pauli matrices.

$$U = \exp\left[i\left(gA_2^\infty z + \frac{\pi}{2}\right)\sigma_2\right] = \cos\left[\frac{1}{2}\left(gA_2^\infty z + \frac{\pi}{2}\right)\right] + i\sigma_2 \sin\left[\frac{1}{2}\left(gA_2^\infty z + \frac{\pi}{2}\right)\right]. \tag{3.43}$$

We then find the transformed gauge field: $A_x \to 0$, $A_y \to 0$, and

$$A_z \to A_z' = U\left(A_2 \frac{\sigma_2}{2}\right)U^\dagger - \frac{i}{g}U\partial_z U^\dagger = (A_2 - A_2^\infty)\frac{\sigma_2}{2} = A_2^\infty \delta\hat{A}_2(r)\frac{\sigma_2}{2} \tag{3.44}$$

so that

$$\mathbf{A}_2 \to \mathbf{A}_2' = (A_2 - A_2^\infty)\hat{z} = A_2^\infty \delta\hat{A}_2 \hat{z} \tag{3.45}$$

as well as

$$A_4 \to A'_4 = U\left(A_{34}\frac{\sigma_3}{2}\right)U^+ - \frac{i}{g}U\partial_4 U^+$$

$$= A_{34}\cos\left(gA_2^\infty z + \frac{\pi}{2}\right)\frac{\sigma_3}{2} - A_{34}\sin\left(gA_2^\infty z + \frac{\pi}{2}\right)\frac{\sigma_1}{2} \qquad (3.46)$$

$$\cong -gA_2^\infty A_{34} z \frac{\sigma_3}{2} - A_{34}\frac{\sigma_1}{2} \ .$$

In the maximal Abelian gauge, the transformed gauge field produces the following chromo-fields near $z = 0$:

$$\mathbf{E}'_1 = \nabla A'_{14} - g\mathbf{A}'_2 A'_{34} \cong -\left(\frac{dA_{34}}{dr}\right)\hat{r}, \qquad (3.47)$$

$$\mathbf{E}'_3 = \nabla A'_{34} + g\mathbf{A}'_2 A'_{14} \cong -gA_2 A_{34}\hat{z}, \qquad (3.48)$$

and

$$\mathbf{B}'_2 = \nabla \times \mathbf{A}'_2 = -\left(\frac{dA_2}{dr}\right)\hat{\phi} \qquad (3.49)$$

so that the transformed gauge field generates a chromo-electric flux tube along the $z$-axis.

Note that $\mathbf{E}'_3$ satisfies

$$\nabla \times \mathbf{E}'_3 \cong \left\{gA_2^\infty \frac{d\left(\delta\hat{A}_2 A_{34}\right)}{dr}\right\}\hat{\phi} \cong -\mathbf{K}'_3, \qquad (3.50)$$

where $\mathbf{K}'_3 = -gA'_{14}\mathbf{B}'_2 + g\mathbf{A}'_2 \times \mathbf{E}'_1$. This relation between $\mathbf{E}'_3$ and $-\mathbf{K}'_3$ is reminiscent of the relation, $\nabla \times \mathbf{E}_A = \mathbf{k}$, observed in lattice simulations in the maximal Abelian gauge for the Abelian projected chromo-electric field $\mathbf{E}_A$ along a chromo-electric flux tube and the magnetic monopole current density $\mathbf{k}$ [2]. This suggests that the chromo-electric flux tubes generated by the transformed gauge field configurations may contribute to the chromo-electric flux tube between two static quarks observed in lattice simulations in the maximal Abelian gauge.

## 8. The Casimir scaling for the string tension

When regulated on a lattice, the gauge field configurations presented in this section generate chromo-electric flux tubes in the three-dimensional space so that each of them carries an energy that grows with the length of the flux tube. If we assume that these configurations significantly contribute to a chromo-electric flux tube between two static color charges, then we can show the Casimir scaling for the string tension $\sigma_r$ for two static color charges in representation $r$ and separated by a distance $R$, which must be chosen not to be too long so that $\sigma_r$ takes a well-defined constant value.

As the energy is carried by both the chromo-electric and the chromo-magnetic fields, the string tension $\sigma_r$ can be defined as

$$\sigma_r \equiv \frac{1}{\text{Tr}[\mathbf{1}_r]} \frac{1}{R} \int_{-R/2}^{R/2} dz' \int dx dy \left\langle \text{Tr}\left[\frac{1}{2}\left\{ (\mathbf{E}_a t_r^a) \cdot (\mathbf{E}_b t_r^b) + (\mathbf{B}_a t_r^a) \cdot (\mathbf{B}_b t_r^b) \right\}\right]\right\rangle, \qquad (3.51)$$

where $\langle \ \rangle$ is a statistical average over all gauge field configurations and the division by $\text{Tr}[\mathbf{1}_r] = d(r)$ yields the string tension per color. $\sigma_r$ then satisfies the following Casimir scaling:

$$\sigma_r = \frac{C_2(r)}{C_2(N)} \sigma_N, \qquad (3.52)$$

where $C_2(r)$ is the quadratic Casimir operator for representation $r$ and $N$ stands for the fundamental representation.

$$\begin{aligned}
\sigma_r &= \frac{1}{\text{Tr}[\mathbf{1}_r]}\frac{1}{R}\int_{-R/2}^{R/2}dz'\int dxdy\left\langle \text{Tr}\left[\frac{1}{2}\{-(\mathbf{E}_a t_r^a)\cdot(\mathbf{E}_b t_r^b)+(\mathbf{B}_a t_r^a)\cdot(\mathbf{B}_b t_r^b)\}\right]\right\rangle \\
&= \frac{1}{d(r)}\frac{1}{R}\int_{-R/2}^{R/2}dz'\int dxdy\,\text{Tr}[t_r^a t_r^b]\left\langle \frac{1}{2}(-\mathbf{E}_a\cdot\mathbf{E}_b+\mathbf{B}_a\cdot\mathbf{B}_b)\right\rangle \\
&= \frac{1}{d(r)}\frac{1}{R}\int_{-R/2}^{R/2}dz'\int dxdy\,C(r)\delta^{ab}\left\langle \frac{1}{2}(-\mathbf{E}_a\cdot\mathbf{E}_b+\mathbf{B}_a\cdot\mathbf{B}_b)\right\rangle \\
&= \frac{C(r)d(N)}{d(r)C(N)}\left[\frac{1}{d(N)}\frac{1}{R}\int_{-R/2}^{R/2}dz'\int dxdy\,C(N)\delta^{ab}\left\langle \frac{1}{2}(-\mathbf{E}_a\cdot\mathbf{E}_b+\mathbf{B}_a\cdot\mathbf{B}_b)\right\rangle\right] \\
&= \frac{C_2(r)d(G)}{d(G)C_2(N)}\left[\frac{1}{\text{Tr}[\mathbf{1}_N]}\frac{1}{R}\int_{-R/2}^{R/2}dz'\int dxdy\left\langle \text{Tr}\left[\frac{1}{2}\{-(\mathbf{E}_a t_r^a)\cdot(\mathbf{E}_b t_r^b)+(\mathbf{B}_a t_r^a)\cdot(\mathbf{B}_b t_r^b)\}\right]\right\rangle\right] \\
&= \frac{C_2(r)}{C_2(N)}\sigma_N,
\end{aligned}$$

(3.53)

where we have used

$$\text{Tr}[t_r^a t_r^b] = C(r)\delta^{ab} \tag{3.54}$$

and

$$\frac{C(r)}{d(r)} = \frac{C_2(r)}{d(G)} \tag{3.55}$$

### C. The gauge configuration with a chromo-magnetic flux tube

#### 1. The gauge field configuration and the chromo-fields

Consider a gauge field configuration that satisfies

$$\mathbf{A}_2 = A_2(r)\hat{z} \tag{3.56}$$

and

$$\mathbf{A}_3 = A_3(r)\hat{\phi}, \tag{3.57}$$

where we use the cylindrical coordinate system, where at each spatial location $\mathbf{r} = (r, \phi, z)$, we have three unit vectors, $\hat{r}$, $\hat{\phi}$, and $\hat{z}$. All the other components of the gauge field are zero. This gauge field configuration satisfies both $\nabla \cdot \mathbf{A}_a = 0$ and $\partial_4 A_{a4} = 0$ so that it satisfies the Landau gauge condition, $\nabla \cdot \mathbf{A}_a + \partial_4 A_{a4} = 0$.

We then find

$$\mathbf{B}_1 = -gC_{123}\mathbf{A}_2 \times \mathbf{A}_3 = gC_{123}A_2(r)A_3(r)\hat{r}, \tag{3.58}$$

$$\mathbf{B}_2 = \nabla \times \mathbf{A}_2 = -\frac{dA_2}{dr}\hat{\phi}, \tag{3.59}$$

and

$$\mathbf{B}_3 = \nabla \times \mathbf{A}_3 = \frac{1}{r}\frac{d}{dr}(rA_3)\hat{z}, \tag{3.60}$$

while $\mathbf{E}_1 = \mathbf{E}_2 = \mathbf{E}_3 = 0$. We will show that these magnetic fields decay exponentially as $r$ is increased so that they represent a chromo-magnetic flux sheet whose three-dimensional cross section is a chromo-magnetic flux tube parallel to the $z$-axis.

When rotated in a plane that includes the temporal $x_4$-axis and is perpendicular to the flux tube axis, each of these configurations acquires non-zero chromo-electric field along the flux tube axis so that the rotated gauge field configuration generates a chromo-electric flux tube. For example, when rotated in the $x - x_4$ plane, $\mathbf{B}_2$ generates chromo-electric field $\mathbf{E}'_2$ along the $z$-axis,

$$\mathbf{E}'_2 \propto \frac{dA_2}{dr}\hat{z} \tag{3.61}$$

so that the rotated gauge field configuration carries a chromo-electric flux along the *z*-axis. As $K_{24} = 0$ and $\mathbf{K}_2 = 0$, the rotated gauge field configuration satisfies $\nabla \times \mathbf{E}'_2 + \partial_4 \mathbf{B}'_2 = 0$ as found in the lattice simulations in the Landau gauge [3]. This suggests that the chromo-electric flux tubes generated by the rotated gauge field configurations may contribute to the chromo-electric flux tube between two static quarks observed in lattice simulations in the Landau gauge.

## 2. The equations for the gauge field

The equation of motion for $\mathbf{B}_2$, $\nabla \times \mathbf{B}_2 = -\mathbf{J}_2$, together with

$$-\mathbf{J}_2 = gC_{231}(\mathbf{A}_3 \times \mathbf{B}_1) = -(gC_{123}A_3)^2 \mathbf{A}_2 \tag{3.62}$$

becomes the following equation for $A_2$

$$\nabla^2 \mathbf{A}_2 = -\nabla \times \nabla \times \mathbf{A}_2 = -\nabla \times \mathbf{B}_2 = \mathbf{J}_2 = (gC_{123}A_3)^2 \mathbf{A}_2 \tag{3.64}$$

so that

$$\left(\frac{d^2}{dr^2} + \frac{1}{r}\frac{\partial}{\partial r}\right)A_2 = (gC_{123}A_3)^2 A_2. \tag{3.65}$$

The equation of motion for $\mathbf{B}_3$, $\nabla \times \mathbf{B}_3 = -\mathbf{J}_3$, together with

$$-\mathbf{J}_3 = gC_{321}(\mathbf{A}_2 \times \mathbf{B}_1) = -(gC_{123}A_2)^2 \mathbf{A}_3 \tag{3.66}$$

becomes the following equation for $A_3$

$$\nabla^2 \mathbf{A}_3 = -\nabla \times \nabla \times \mathbf{A}_3 = -\nabla \times \mathbf{B}_3 = \mathbf{J}_3 = (gC_{123}A_2)^2 \mathbf{A}_3 \tag{3.67}$$

so that

$$\left(\frac{d^2}{dr^2} + \frac{1}{r}\frac{\partial}{\partial r} - \frac{1}{r^2}\right)A_3 = (gC_{123}A_2)^2 A_3. \tag{3.68}$$

### 3. The solution for the gauge field away from the core of a chromo-field flux tube

To solve the above equations for $A_2$ and $A_3$, we impose the following asymptotic boundary conditions:

$$\lim_{r \to \infty} A_3 = 0 \tag{3.69}$$

and

$$\lim_{r \to \infty} A_2 = \text{const} = A_2^\infty. \tag{3.70}$$

For dimensionless variables, $\hat{A}_3$ and $\delta \hat{A}_2$, defined by

$$A_3 \equiv |A_2^\infty| \hat{A}_3 \tag{3.71}$$

and

$$A_2 \equiv A_2^\infty \left(1 + \delta \hat{A}_2 \right), \tag{3.72}$$

the above equations become

$$\left( \frac{d^2}{d\rho^2} + \frac{1}{\rho} \frac{d}{d\rho} - \frac{1}{\rho^2} \right) \hat{A}_3 = \left(1 + \delta \hat{A}_2 \right)^2 \hat{A}_3 \tag{3.73}$$

and

$$\left( \frac{d^2}{d\rho^2} + \frac{1}{\rho} \frac{d}{d\rho} \right) \delta \hat{A}_2 = \left( \hat{A}_3 \right)^2 \left(1 + \delta \hat{A}_2 \right), \tag{3.74}$$

where the scaled length radial distance $\rho$ is defined by $\rho \equiv g C_{123} |A_2^\infty| r$.

Assuming $|\delta \hat{A}_2| \ll 1$ for $\rho \gg 1$, which we will justify later, we can approximate these equation for $\hat{A}_3$ and $\delta \hat{A}_2$ by

$$\left( \frac{d^2}{d\rho^2} + \frac{1}{\rho} \frac{d}{d\rho} - \frac{1}{\rho^2} \right) \hat{A}_3 \cong \hat{A}_3 \tag{3.75}$$

and

$$\left(\frac{d^2}{d\rho^2}+\frac{1}{\rho}\frac{d}{d\rho}\right)\delta\hat{A}_2 \cong \left(\hat{A}_3\right)^2. \tag{3.76}$$

The solution of the equation for $\hat{A}_3$ is then

$$\hat{A}_3 \cong c_\phi K_1(\rho), \tag{3.77}$$

where $K_1$ is the modified Bessel function of the second kind and $c_\phi$ is a constant that sets the scale for $\hat{A}_3$. We then obtain

$$A_3 \cong c_\phi |A_2^\infty| K_1\left(-gC_{123}|A_2^\infty|r\right), \tag{3.78}$$

The equation for $\delta\hat{A}_2$ then becomes

$$\frac{1}{\rho}\frac{d}{d\rho}\left(\rho\frac{d}{d\rho}\right)\delta\hat{A}_2 \cong \left(c_\phi K_1\right)^2. \tag{3.79}$$

Since for $\rho \gg 1$,

$$K_1(\rho) \cong \sqrt{\frac{\pi}{2\rho}}\exp(-\rho), \tag{3.80}$$

following the steps similar to those in Sec.III.B, we can obtain

$$A_2 = A_2^\infty\left\{1-\left(\frac{\pi c_\phi^2}{4}\right)\text{Ei}(-2\rho)\right\}, \tag{3.81}$$

where Ei is the exponential integral.

For $r \gg 1$, we then find both $\hat{A}_3$ and $\delta\hat{A}_2$ to decay exponentially so that

$$A_3 \cong \sqrt{\frac{\pi}{2}} c_\phi |A_2^\infty| \sqrt{\frac{\lambda}{r}} \exp\left(-\frac{r}{\lambda}\right), \tag{3.82}$$

where $\lambda$ is the decay length for $\hat{A}_{34}$, and

$$A_2 \cong A_2^\infty \left\{ 1 + \left(\frac{\pi c_4^2}{4}\right) \frac{\xi}{r} \exp\left(-\frac{r}{\xi}\right) \right\}, \tag{3.83}$$

where $\xi$ is the decay length for $\delta\hat{A}_2$.

Note that $A_2$ is somewhat analogous to the Cooper pair condensate in the Ginzburg-Landau theory of superconductors while $\mathbf{A}_3$ is analogous to the vector potential for a magnetic field in a superconductor. $\lambda$ is then analogous to the penetration length while $\xi$ is analogous to the coherence length. As $\lambda/\xi = 2 > 1/\sqrt{2}$, the chromo-magnetic flux tube is analogous to a magnetic vortex in a type II superconductor.

### 4. The squeezing of the chromo-electric and magnetic fields

For $r \gg 1$, we can then show that the chromo-magnetic fields, $\mathbf{B}_1$ and $\mathbf{B}_3$, decay exponentially with the decay length $\lambda$ for $\hat{A}_3$ while the chromo-magnetic field $\mathbf{B}_2$ decays exponentially with the decay length $\xi$ for $\delta\hat{A}_2$:

$$\mathbf{B}_1 = -c_\phi g C_{123} (A_2^\infty)^2 K_1\left(-\frac{r}{\lambda}\right) \hat{r} \cong -\sqrt{\frac{\pi}{2}} c_\phi g C_{123} (A_2^\infty)^2 \sqrt{\frac{\lambda}{r}} \exp\left(-\frac{r}{\lambda}\right) \hat{r} \tag{3.84}$$

$$\begin{aligned} \mathbf{B}_3 &\cong c_\phi \left[ g C_{123} (A_2^\infty)^2 \frac{1}{2}\left\{ K_0\left(-\frac{r}{\lambda}\right) + K_2\left(-\frac{r}{\lambda}\right) \right\} + \frac{1}{r} A_2^\infty K_1\left(-\frac{r}{\lambda}\right) \right] \hat{z} \\ &\cong \sqrt{\frac{\pi}{2}} g C_{123} (A_2^\infty)^2 \sqrt{\frac{r}{\lambda}} \exp\left(-\frac{r}{\lambda}\right) \hat{z}, \end{aligned} \tag{3.85}$$

and

$$\mathbf{B}_2 \cong \left(\frac{\pi c_\phi^{\ 2}}{4}\right) \frac{A_2^\infty}{r} \exp\left(-\frac{r}{\xi}\right) \hat{\phi}. \qquad (3.86)$$

We can also show that far away from the *z*-axis so that $\mathbf{A}_2 \cong A_2^\infty \hat{z}$ and $\mathbf{B}_2 \cong 0$, the chromomagnetic field $\mathbf{B}_3$ satisfies

$$\nabla^2 \mathbf{B}_3 \cong \left(gC_{123}A_2^\infty\right)^2 \mathbf{B}_3, \qquad (3.87)$$

which is analogous to the equation describing the Meissner effect for the magnetic field in a superconductor. In fact, this equation follows from $\nabla \times \mathbf{B}_3 = -\mathbf{J}_3$, which is analogous to the equation for Ampere's law for the magnetic field in the superconductor, and $-\mathbf{J}_3 \cong -\left(gC_{123}A_2^\infty\right)^2 \mathbf{A}_3$, which is analogous to the London equation for the superconductor.

$$\nabla^2 \mathbf{B}_3 = \nabla(\nabla \cdot \mathbf{B}_3) - \nabla \times \nabla \times \mathbf{B}_3 = -\nabla \times (-\mathbf{J}_3) \cong \left(gC_{123}A_2^\infty\right)^2 (\nabla \times \mathbf{A}_3) = \left(gC_{123}A_2^\infty\right)^2 \mathbf{B}_3.$$

$$(3.88)$$

Similarly, we find

$$\nabla^2 \mathbf{B}_1 \cong \left(gC_{123}A_2^\infty\right)^2 \mathbf{B}_1, \qquad (3.89)$$

which follows from $\nabla \cdot \mathbf{B}_1 = K_{14}$, where $K_{14} \cong gC_{123}\mathbf{A}_2 \cdot \mathbf{B}_3$, and $\nabla \times \mathbf{B}_3 = -\mathbf{J}_3$, where $-\mathbf{J}_3 \cong -\left(gC_{123}A_2^\infty\right)^2 \mathbf{A}_3$:

$$\begin{aligned}
\nabla^2 \mathbf{B}_1 &= \nabla(\nabla \cdot \mathbf{B}_1) - \nabla \times \nabla \times \mathbf{B}_1 = \nabla K_{14} \cong gC_{123}\nabla(\mathbf{A}_2 \cdot \mathbf{B}_3) \\
&= gC_{123}\left\{\mathbf{A}_2 \times (\nabla \times \mathbf{B}_3) + \mathbf{B}_3 \times (\nabla \times \mathbf{A}_2) + (\mathbf{A}_2 \cdot \nabla)\mathbf{B}_3 + (\mathbf{B}_3 \cdot \nabla)\mathbf{A}_2\right\} \quad (3.90) \\
&\cong gC_{123}\left\{\mathbf{A}_2 \times (-\mathbf{J}_3)\right\} = -\left(gC_{123}A_2^\infty\right)^2 gC_{123}(\mathbf{A}_2 \times \mathbf{A}_3) = \left(gC_{123}A_2^\infty\right)^2 \mathbf{B}_1.
\end{aligned}$$

## 5. The gauge field diverges logarithmically at the core of a chromo-magnetic flux tube

We can show that at the core of the chromo-magnetic flux tube, the non-zero components of the gauge field satisfying Eq.(3.65) and Eq.(3.68) diverge logarithmically. We can regularize these divergences by approximating the path integral for the partition function by the corresponding lattice gauge theory on a four-dimensional lattice.

As the non-zero components of the gauge field tend to diverge near the center of a chromo-magnetic flux tube, the link variables corresponding to these gauge field components start to sample non-trivial center elements in the gauge group so that in lattice simulations, these chromo-magnetic flux tubes may be detected as center vortices.

We can show that near the core of the chromo-electric flux tube, the chromo-magnetic fields all diverge logarithmically. We can then regularize these divergences by approximating the path integral for the partition function by the corresponding lattice gauge theory on a four-dimensional lattice.

As the non-zero components of the gauge fields tend to diverge near the center of a chromo-field flux tube, the link variables corresponding to these gauge field components start to sample non-trivial center elements in the gauge group so that in lattice simulations, these chromo-field flux tubes may be detected as center vortices.

## 6. The Euclidean action for the gauge field configuration

As $\mathbf{B}_2 \cdot \mathbf{B}_2 \propto c_\phi^{\,4} (A_2^\infty)^2$ and $\mathbf{B}_1 \cdot \mathbf{B}_1 + \mathbf{B}_3 \cdot \mathbf{B}_3 \propto c_\phi^{\,2} (A_2^\infty)^3$, the Euclidean action $S_E$ for the lattice gauge field configuration corresponding to our gauge field configuration is also controlled by $c_4$ and $A_2^\infty$. Particularly, the value of $S_E$ can be arbitrarily small so that these lattice gauge field configurations should contribute significantly to the vacuum state of the Yang-Mills theory with high statistical weights as the statistical weight or probability for each configuration is proportional to $\exp(-S_E)$. This is why we believe that these gauge field configurations must significantly contribute to the low-energy properties of the SU(N) Yang Mills theory, particularly QCD.

### D. The Gribov horizon

When regulated on a lattice, a gauge field configuration is found to be located on the Gribov horizon, the boundary of the Gribov region, if the following Faddeev-Popov operator has a non-trivial zero eigenvalue:

$$M_{\mathbf{rr'}}^{ab} \equiv \delta^{ab}\sum_i \left(2\delta_{\mathbf{rr'}} - \delta_{\mathbf{r}+\hat{i},\mathbf{r'}} - \delta_{\mathbf{r}-\hat{i},\mathbf{r'}}\right) + \frac{1}{2}gC_{acb}\sum_i \left[-A_{ci}(\mathbf{r})\delta_{\mathbf{r}+\hat{i},\mathbf{r'}} + A_{ci}(\mathbf{r'})\delta_{\mathbf{r}-\hat{i},\mathbf{r'}}\right], \quad (3.91)$$

where $\mathbf{r}$ and $\mathbf{r'}$ denote lattice sites and we have omitted a term in the operator that vanishes at the continuum limit [9]. In the following, we will argue that when regulated on a lattice with a volume $V = L^4$, a flux sheet gauge field configuration belonging to one of the two sets presented in Sec.III.B and Sec.III.C is located on the Gribov horizon as long as the thickness, $\lambda = 1/(gC_{123}|A_2^\infty|)$, of its flux sheet is much shorter than $L$ (i.e., $\lambda \ll L$), which implies that in the infinite lattice volume limit, any of these flux sheet gauge field configurations with a finite flux sheet thickness is located on the Gribov horizon.

As shown in Sec.III.B.3 and Sec.III.C.3, these two types of gauge field configurations are built from an off-diagonal component $\mathbf{A}_2 = A_2^\infty (1 + \delta \hat{A}_2)\hat{z}$ whose magnitude becomes almost constant at $A_2^\infty$ outside a flux sheet or for $r \gg \lambda = 2\xi$, and a diagonal component, $A_{34} = |A_2^\infty|\hat{A}_{34}$ or $\mathbf{A}_3 = |A_2^\infty|\hat{A}_3\hat{\phi}$, whose magnitude practically vanishes outside the flux sheet or for $r \gg \lambda$. We can then express the second term in the Faddeev-Popov operator as

$$\frac{1}{2}gC_{acb}\sum_i \left[-A_{ci}(\mathbf{r})\delta_{\mathbf{r}+\hat{i},\mathbf{r'}} + A_{ci}(\mathbf{r'})\delta_{\mathbf{r}-\hat{i},\mathbf{r'}}\right] = \frac{1}{2}gC_{a2b}A_2^\infty\left(-\delta_{\mathbf{r}+\hat{z},\mathbf{r'}} + \delta_{\mathbf{r}-\hat{z},\mathbf{r'}}\right) + \delta K_{\mathbf{rr'}}^{ab}, \quad (3.92)$$

where the operator $\delta K^{ab}_{\mathbf{rr}'}$ consists of terms controlled by $\delta \hat{A}_2$ as well as by $\hat{A}_{34}$ or $\hat{A}_3$ so that it practically vanishes outside the flux sheet as long as $\lambda \ll L$ is satisfied. We will then treat $\delta K^{ab}_{\mathbf{rr}'}$ as a perturbation to

$$K^{ab}_{\mathbf{rr}'} = \delta^{ab} \sum_i \left( 2\delta_{\mathbf{rr}'} - \delta_{\mathbf{r}+\hat{i},\mathbf{r}'} - \delta_{\mathbf{r}-\hat{i},\mathbf{r}'} \right) + \frac{1}{2} g C_{a2b} A_2^{\infty} \left( -\delta_{\mathbf{r}+\hat{z},\mathbf{r}'} + \delta_{\mathbf{r}-\hat{z},\mathbf{r}'} \right). \quad (3.93)$$

We can then show that the following function,

$$\phi^a_{\mathbf{r},n\pm} \equiv \frac{1}{\sqrt{V}} \exp(ipz) \frac{\delta^{1a} \pm i\delta^{3a}}{\sqrt{2}}, \quad (3.94)$$

where $p = 2\pi n/L$ with $n$ being an integer in an interval $-L/2 < n \leq L/2$), satisfies the following eigenvalue equation,

$$\sum_{\mathbf{r}',b} K^{ab}_{\mathbf{rr}'} \phi^b_{\mathbf{r}',n\pm} = \eta_{n\pm} \phi^a_{\mathbf{r},n\pm}, \quad (3.95)$$

where the eigenvalue $\eta_{n\pm}$ is given by

$$\eta_{n\pm} = 2(1 - \cos p) \pm g C_{123} A_2^{\infty} \sin p. \quad (3.96)$$

We can show that there exists a non-zero value of $p$, for which this eigenvalue vanishes so that gauge field configurations satisfying $\mathbf{A}_2 = A_2^{\infty} \hat{z}$ as well as $A_{34} = 0$ or $\mathbf{A}_3 = 0$, admit a non-trivial eigenvalue for the Faddeev-Popov operator and they are therefore on the Gribov horizon.

Treating $\delta K^{ab}_{\mathbf{rr}'}$ as a perturbation to $K^{ab}_{\mathbf{rr}'}$, we can then obtain a correction to the eigenvalue $\eta_{n\pm}$. As $\delta K^{ab}_{\mathbf{rr}'}$ practically vanishes outside the flux sheet, the first-order correction should scale as

$$\Delta \eta_{n\pm}^{(1)} = \langle \phi_{\mathbf{r},n\pm}^{a} | \delta K | \phi_{\mathbf{r},n\pm}^{a} \rangle \sim \lambda^2 L^2 / V = (\lambda / L)^2, \qquad (3.97)$$

where $V$ comes from $|\phi_{\mathbf{r},n\pm}^{1}|^2 + |\phi_{\mathbf{r},n\pm}^{3}|^2 = 1/V$, while the second-order correction should scale with $(\lambda/L)^4 L \propto (\lambda/L)^3$, where $L$ next to $(\lambda/L)^4$ follows from the spacing, $\Delta p = 2\pi/L$, between two successive values for $p$. In general, the $k$-th order correction should scale with $(\lambda/L)^{2k} L^{k-1} \propto (\lambda/L)^{k+1}$. As long as $\lambda \ll L$ is satisfied for the flux sheet gauge field configuration, the correction to $\eta_{n\pm}$ must be negligible so that this gauge field configuration admits a non-trivial zero eigenvalue for the Faddeev-Popov operator and it is therefore located on the Gribov horizon.

## IV. Conclusions and outlook

For the four-dimensional SU(N) Euclidean Yang-Mills theory in the Landau gauge, we have presented two sets of gauge field configurations that satisfy the Euclidean equations of motion. These configurations generate four-dimensional chromo-field flux sheets whose spatial cross sections are three-dimensional chromo-field flux tubes. In lattice simulations, they may be detected as center vortices.

The first set of gauge field configurations generates chromo-electric flux tubes that should contribute to a chromo-electric flux tube between two static color charges. The string tension $\sigma_r$ for two static color charges in representation $r$ then naturally satisfies the Casimir scaling. Applying a gauge transformation to this set of gauge field configurations, we can transform them into those in the maximal Abelian gauge. These transformed configurations generate chromo-electric flux tubes that should contribute to those observed between two static quarks in lattice simulations performed in the maximal Abelian gauge.

The second set of gauge field configurations generates chromo-magnetic flux tubes. When rotated in a plane that includes the temporal $x_4$-axis and is perpendicular to the flux tube axis, the rotated gauge field configuration generates a chromo-electric flux tube and should contribute to the chromo-electric flux tubes observed in lattice simulations in the Landau gauge.

We have also argued that when regulated on a lattice with a volume $V = L^4$, a flux sheet gauge field configuration belonging to one of these two sets is located on the Gribov horizon as long as the thickness $\lambda$ of its flux sheet is much shorter than $L$ (*i.e.*, $\lambda << L$) so that in the infinite lattice volume limit, any flux sheet gauge field configuration with a finite flux sheet thickness is located on the Gribov horizon.

We thus suggest that these configurations contribute significantly to the low energy properties of QCD, particularly the quark confinement. We may also suggest that these gauge field configurations disorder the vacuum state of the SU(N) Euclidean Yang Mills theory producing the area law for the expectation value for the Wilson loop and generating a finite correlation length or a mass gap for the Euclidean correlation functions for gauge invariant quantities.

Questions that were not addressed in this article but should be examined in the future are (1) whether these gauge field configurations are responsible for the chiral symmetry breaking, another important low energy property of QCD; (2) how these configurations affect the properties of QCD at finite temperatures, especially the deconfinement transition; (3) how these configurations are related to the gluon propagator obtained from the Schwinger-Dyson equations as well as in lattice simulations [10].

## Acknowledgments

I wish to thank Michele Bock for constant support and encouragement and Richard F. Martin, Jr. and other members of the physics department at Illinois State University for creating a supportive academic environment. I also wish to thank the reviewer of this article for suggesting to check whether the gauge field configurations presented in this article are located on the Gribov horizon.